\documentstyle[12pt]{article}

\begin{document}

\begin{titlepage}

\begin{tabular}{l}
October, 1994                   
\end{tabular}
    \hfill
\begin{tabular}{l}
MSU-HEP-41024 \\
CTEQ-404 \\
\end{tabular}

\vspace{2cm}

\begin{center}

\renewcommand{\thefootnote}{\fnsymbol{footnote}}
{
\LARGE Global QCD Analysis and \\the CTEQ Parton Distributions%
\footnote[2]{This work is supported in part by DOE, NSF and TNRLC}
}
\renewcommand{\thefootnote}{\arabic{footnote}}

\vspace{1.25cm}

{\large
H.L. Lai$^a$, J. Botts$^{a,b}$, J. Huston$^a$, J.G. Morfin$^c$,  \\
J.F. Owens$^d$, J.W. Qiu$^e$, W.K. Tung$^a$ and H. Weerts$^a$
}

\vspace{1.25cm}

$^a$Michigan State University, $^b$DESY-Zeuthen, \\
$^c$Fermi National Accelerator Laboratory, \\
$^d$Florida State University, $^e$Iowa State University

\end{center}
\vfill

\begin{abstract}
The CTEQ program for the determination of parton distributions through a
global QCD analysis of data for various hard scattering processes is fully
described. A new set of distributions, CTEQ3, incorporating several new
types of data is reported and compared to the two previous sets of CTEQ
distributions. Comparison with current data is discussed in some detail. The
remaining uncertainties in the parton distributions and methods to further
reduce them are assessed. Comparisons with the results of other global
analyses are also presented.
\end{abstract}

\vfill
\newpage
\end{titlepage}

\section{Introduction\label{sec:intro}}

Calculations of high energy hard scattering processes in perturbative
quantum chromodynamics (pQCD) rely on two basic ingredients -- (1) the
perturbatively calculated scattering cross-sections involving the
fundamental partons, leptons, and gauge bosons and (2) the parton
distributions inside the incoming hadrons. Our knowledge of these universal
parton distributions functions (PDF's) is derived, in turn, from the
analysis of data for a variety of hard scattering processes. Early analyses
were often limited to deep inelastic lepton nucleon scattering and lepton
pair production, as these were the processes for which extensive data sets
were available and for which next-to-leading-order (NLO) calculations of the
hard scattering subprocesses had been performed. Now the number of available
NLO calculations has increased and, simultaneously, data for additional hard
scattering processes have become available for a variety of beams and
targets. This progress makes it possible to determine the parton
distributions with a greater precision than was previously possible. Indeed,
assumptions such as an SU(3) or SU(2) symmetry for the quark-antiquark sea
in the proton have had to be discarded in the face of experimental evidence
to the contrary.

With this wealth of data and corresponding theoretical calculations, true
``global analyses'' have become possible. In such a program there are two
main goals. The first is to determine the parton distributions as precisely
as possible, using all available experimental input, and to suggest which
new types of data are necessary in order to further improve the precision. A
review of progress in this area and references to earlier work can be found
in Ref.\cite{OwTu}. Secondly, with an over-constrained set of PDF's it
becomes possible to explore whether or not the parton-level calculations in
pQCD constitute a consistent theoretical framework to account for all the
available experimental data relevant for pQCD studies. This may point to
areas where improved theoretical treatments are required and, perhaps,
uncover areas where various data sets used in the analysis might be mutually
inconsistent. Either way, one can expect important progress to be made as
the result of careful and critical confrontation of data with theory. This
potential has been discussed in some length in Ref.\cite{Tung94a}.

This paper describes the series of global analyses conducted by the CTEQ
Collaboration.\footnote{%
The CTEQ collaboration (Coordinated Theoretical/Experimental Project on QCD
Phenomenology and Tests of the Standard Model) consists of, in addition to
the above authors (as members of its global fit subgroup), S. Kuhlmann
(ANL); S. Mishra (Harvard); R. Brock, J. Pumplin, C.P. Yuan (MSU); D. Soper
(Oregon); J. Collins, J. Whitmore (PSU); F. Olness (SMU); and J. Smith, G.
Sterman (Stony Brook).} The necessary tools for carrying out this systematic
global analysis program have been developed from those used in the previous
work of Duke, Owens, and collaborators \cite{DuOw}, \cite{DDOR} and
Morfin-Tung \cite{MoTu}. The use of two independent QCD parton evolution and
global fitting programs provides a valuable consistency check on all aspects
of the analysis. The CTEQ program is designed to systematically refine the
PDF's as new theoretical and experimental advances are made, and to clearly
describe the theoretical and experimental inputs and their relation to the
resulting parton distributions. The different versions of CTEQ distributions
reflect historically a series of different assumptions and inputs. As a
general rule, newer versions incorporate more up-to-date data and are
preferred {\em overall} than earlier ones, although this may not be an
absolute statement because the multi-faceted nature of global analysis does
not always lead to a one-dimensional progression of improvements---as will
become clear when these developments are described. It should be recognized
that, in general, differences between current and prior PDF's are not a
reflection of ``theoretical uncertainties'', but rather are indications of
the manner in which new developments in data/theory impact on the
determination the underlying parton distributions.

The first stage of this analysis (known as ``CTEQ1''\cite{CTEQ1}), was
performed in 1992 following the availability of the high precision deep
inelastic scattering (DIS) data by the CCFR \cite{CCFR} and NMC \cite{NMC-N}
collaborations. A second, unpublished set (known as ``CTEQ2''), spurred by
new data from HERA \cite{HERA93}, has been circulated during the past year.%
\footnote{%
A brief description of these distributions has been given in Ref.\cite
{Tung94a}.} The advent of recent data on the lepton asymmetry in $W-$boson
production \cite{CDF-W} and on the difference in Drell-Yan cross-sections
from proton and neutron targets \cite{NA51} has stimulated further
refinements which result in a new set which will be referred to as
``CTEQ3.'' The common features as well as differences amongst these three
sets will be discussed in detail in this paper.

In Secs.~\ref{sec:expt} and \ref{sec:analysis} below, we review the various
physical processes and experimental inputs included in our analysis and
present a relatively self-contained account of the analysis and fitting
procedures used. The development of the three versions of CTEQ distributions
is described in Sec.~\ref{sec:PDFs}, reserving the most detailed discussion
to the latest version. Comparisons with other parton distributions and with
recent data are described in Sec.~\ref{sec:compare}. Some remaining
uncertainties in the parton distributions and outstanding challenges are
discussed in Sec.~\ref{sec:uncertainty}. Our conclusions are given in Sec.~%
\ref{sec:summary}. Readers with immediate interest in results and recent
developments can skip to Sec.~\ref{sec:PDFs} and refer back for necessary
details.

A similar program of global analyses and continual upgrading of PDF's has
been undertaken by the MRS group \cite{MRSD,MRSA}. The recently released
MRS(H) distributions have now been revised and replaced by the MRS(A) set as
a result of new data mentioned above.\footnote{%
It is worth noting that the MRS analyses are based on the evolution and
fitting package developed jointly by Duke, Owens, and Roberts some time ago
\cite{DuOw}, \cite{DDOR}, \cite{DOR}; hence the tools of analysis of the MRS
and CTEQ groups, in fact, overlap. There are, however, differences in
analysis procedures, data selection, and the handling of experimental errors.%
} A comparison of these results with those of our analysis is included in
Sec.\ref{sec:compare}.

\section{Experimental Input\label{sec:expt}}

In order to make the comparison of theory with experiment well-defined, we
have limited the kinematic range to that where the ``leading twist'' QCD
contributions are dominant. Thus, we restrict the selection of experimental
data to kinematic regions which contain at least one large momentum scale ``$%
Q$'' $>Q_{min}$. In the absence of a reliable theoretical guide in the
perturbative formalism, the value $Q_{min}$ is varied within the range $2-10$
GeV to test for sensitivity of the results to this choice. We found stable
results generally with the following minimum kinematic constraints: for deep
inelastic scattering, $Q$ (virtuality of the vector boson) $>2$ GeV and $W$
(CM energy) $>3.5$ GeV; for lepton-pair production, $Q>4$ GeV; for direct
photon production $p_T>4$ GeV.

Recent high statistics DIS data from NMC \cite{NMC-N} on $%
F_2^n/F_2^p,F_2^p-F_2^n,$ and $F_2^{p,d}$ using a muon beam and from CCFR
\cite{CCFR} on $F_{2,3}^{Fe}$ using (anti-) neutrinos, combined with the
existing BCDMS \cite{BCDMS} results provide excellent coverage of the
kinematic region $x>0.01$. New measurements of $F_2^p$ from HERA \cite
{HERA93} have extended the kinematic range down to very small $x$ values,
approaching $10^{-4}$. Although the errors are comparatively large, the
extended range provides useful constraints on the behavior of the parton
distributions in the small-$x$ region. (Throughout this paper, ``small-$x$''
means $10^{-4}<x<10^{-2}$.) As will be discussed in some detail in Sec.~\ref
{sec:PDFs}, this is particularly important in light of questions raised
concerning the consistency of the structure functions measured in the other
experiments in the intermediate region $0.01<x<0.1$ \cite{CTEQ1}.

Precision data from the SLAC-MIT series of experiments \cite{SlacMit}
largely lie outside the kinematic cuts (especially when the cuts are raised
above the minimum quoted above); and those data points within the cuts agree
well with the BCDMS and NMC data already included. They are thus not used in
the analyses reported here. Data from the earlier EMC experiment \cite
{EMC-F2} are excluded since the disagreement between these data with other
data sets appears to be understood now as the result of the new NMC
analysis. Data from the CDHSW neutrino experiment \cite{CDHSW} are also not
used since in the (wide) region where they agree with the CCFR results, the
latter completely dominate due to the much smaller errors; and in the
(narrow) region where they disagree, it would be inconsistent to include
both sets.\footnote{%
The resolution of this experimental disagreement lies outside the scope of
our work.}

To apply the selected experimental results to the study of the parton
structure of the nucleon, the heavy target neutrino data must be converted
to their nucleon equivalent. This conversion is done using measured light to
heavy target ratios obtained in electron and muon scattering processes by
the SLAC-MIT \cite{SLAC-A}, EMC \cite{EMC-A}, and NMC \cite{NMC-A}
experiments. There is an uncertainty associated with this procedure, which
will be commented on later.

DIS data by themselves are not sufficient to provide a complete set of
constraints on the parton content of the nucleon, since the measured nucleon
structure functions represent only a few independent combinations of the
parton flavors. Lepton-pair production experiments provide a useful handle
on the anti-quark distributions (through the $q-\bar q$ annihilation
mechanism) and the gluon distribution (through the $q-g$ ``Compton
scattering'' mechanism). From fixed-target experiments we use the full data
set on the double-differential cross-section $d^2\sigma /d\tau dy$ measured
by the high statistics E605 experiment at Fermilab \cite{E605}. We also
include the new collider data on lepton-pairs measured by the CDF
Collaboration \cite{CDF-lpp}. Although the errors on these data are
comparatively large, they do provide some constraints in the $x$ $\sim
10^{-2}$ region which is beyond the reach of fixed-target experiments.

Another independent source of information is direct photon production which
is particularly sensitive to the gluon distribution. In addition to the
commonly used WA70 data \cite{WA70}, we also include results from the UA6
\cite{UA6} and E706 \cite{E706} experiments. Together these provide coverage
of the region $0.27<x<0.54$ and, hence, help to constrain the gluon
distribution in the middle range of $x$. The deep inelastic data provide
some constraint on the gluon for smaller values of $x$ through the slope of $%
F_2$ with respect to $Q^2$. Additional information at small values of $x$ is
provided by direct photon data from various collider experiments. Indeed,
the coverage in $x$ extends now down to about 0.02 making a simultaneous
analysis of all of the available direct photon data a potentially powerful
tool for extracting the gluon distribution. However, there are still
unresolved theoretical problems associated with understanding the full range
of inclusive (mostly fixed target) and isolated (mostly collider) direct
photon data which need further study. Such a project has been initiated by
members of the CTEQ Collaboration and the results will be presented
elsewhere \cite{CTEQphoton}. For the purpose of the present work only the
fixed target results on inclusive photon cross-section cited above have been
used.

Two new types of data have become available in the past year and they have
provided valuable information on PDF's, notably flavor differentiation of
partons, which were not fully covered by earlier data sets. In particular,
NA51 \cite{NA51} measured the difference of cross sections for producing
lepton pairs at $y=0$ from proton and neutron targets. As discussed in \cite
{ElSt}, this is particularly sensitive to the difference of the $\overline{u}
$ and $\overline{d}$ distributions. And the CDF Collaboration has presented
new data on the charge asymmetry of the decay leptons in $W$ production \cite
{CDF-W}. This measurement contributes to the differentiation of the valence $%
u$ and $d$ quarks as well as the sea-quarks. The effects of including these
two data sets will be discussed further in the next section.

The full data sets we use are summarized in Table \ref{tbl:dataset}.




\section{The CTEQ Global Analysis Program\label{sec:analysis}}

\subsection{Global Analysis Procedures \label{subs:fitting}}

Our goals in the global analysis program are two-fold. On the one hand, we
are seeking a universal set of parton distributions which provide an
accurate description of all of the data sets and are therefore suitable for
use in the calculation of other high energy processes. On the other, we wish
to determine to what degree the theoretical treatment of the hard scattering
processes in the pQCD framework is consistent with all the available
experimental results.

To this end, except where otherwise noted, all data sets included in the
analysis are treated on the same footing. This is to be contrasted with an
often adopted procedure of emphasizing DIS data as the primary source of
information (hence, performing a least-$\chi ^2$ fit to these data alone),
using the other processes only as supplementary constraints. The
simultaneous fitting of many different types of data necessitates the
inclusion of both systematic as well as statistical errors. The systematic
errors include both overall and point-to-point errors. The treatment of the
latter poses a particularly difficult problem. The proper treatment of such
errors typically differs from one experiment to another and doing this for
all experiments requires a prohibitive amount of computer resources. We
studied the impact on the global fit of a full-scale treatment of the
(correlated) systematic errors from the high statistics CCFR and BCDMS
experiments compared to the common practice of combining the point-to-point
systematic and statistical errors in quadrature. The difference is not
significant. Thus, we use the latter procedure as an adequate compromise out
of practical necessity. (Clearly, a fine-tuning of the final results,
including a full treatment of the errors for selected data sets, is possible
if necessary.)

The treatment of the overall normalization errors utilized in CTEQ analyses
differs from that employed by other groups (including most early PDF's, see
\cite{OwTu}, and \cite{MRSD}) which usually allow all experimental data sets
to be varied freely. In our analysis, with the exception of data which
pertain to measured ratios, the normalization (fitting) parameter $N_i$ for
each data set $i$ is associated with a fully correlated error $\epsilon _i$
given by the experiment: a term of the form $(1-N_i)^2/\epsilon _i^2$ is
then added to the overall $\chi ^2$ in the fitting process. This procedure
properly takes into account the normalization uncertainties of the
experiments, whereas the usual practice mentioned above technically
corresponds to assuming infinite normalization errors for all experiments.

The hard cross-sections of all processes included in the analysis are
calculated in pQCD to NLO in $\alpha _s$. We use the
\mbox{$\overline{\rm MS}$} scheme with 5 flavors as the standard,
cf. Sec. \ref
{subs:parametrization} for more details. While such calculations are
generally less sensitive than leading-order (LO) results to the choice of
the renormalization and factorization scales (denoted jointly by the symbol $%
\mu $), the residual dependence on these choices provides a potentially
important source of theoretical uncertainty. In principle, this uncertainty
is one order higher than the approximation used, {\it i.e.},
next-to-next-to-leading order in our case. In practice, it has been learned
that the size of the uncertainty is process-dependent. It is relatively
small for DIS and for lepton-pair production and one usually chooses $\mu =Q$%
, the virtuality of the exchanged virtual vector boson, since this is the
natural large scale in the problem. On the other hand, the NLO predictions
for direct photon production are still sensitive to the choice of $\mu .$ It
is important to address this issue if quantitative results on the gluon
distribution are to be extracted. The common practice of making a specific
choice (say $\mu =p_T)$ without discussion implicitly introduces a bias into
the analysis because of the non-negligible $\mu $-dependence. In this
analysis, we have made the first attempt to address this issue by assigning
a ``theoretical error'' to the predictions associated with the choice of $%
\mu $. The size of this error is estimated by computing the range of
predictions spanned by $\mu =p_T/2$ to $\mu =2p_T$. During the process of
fitting, we let the scale parameter $\mu$ for direct photon calculation
float and add a contribution to the overall $\chi^2$ due to scale uncertainty
given by the deviation of $\mu$ from $p_T$ divided by the ``error''
defined above. Although the details of
this procedure (such as the central value for $\mu$ and the exact
range used to estimate the error) may be
the subject of some debate, it nevertheless represents a reasonable
treatment of the theoretical uncertainty which otherwise is simply
ignored.

\subsection{Relation between PDF's and Observables\label{sec:Relations}}

The relationship between PDF's and the experimental input is in general
quite involved since all parton flavors contribute to the NLO formulas for
the hard cross-section; and, in addition, the parton distribution functions
always mix as the result of QCD evolution. Nonetheless, simple leading order
parton model formulas neglecting small sea-quark contributions are often
useful in providing a qualitative guide to analysis strategies. We will
review the most relevant relations, with the understanding that they are
modified by NLO corrections in practice (to varying degrees for different
processes).

Consider, first, deep inelastic scattering. The available high statistics
data come in four different types, the expressions for which are, in lowest
order, given as follows.

\begin{equation}
\begin{array}{rcl}
F_2^{\mu p} & = & x[4(u+\overline{u})+(d+\overline{d})+2s]/9 \\
F_2^{\mu n} & = & x[4(d+\overline{d})+(u+\overline{u})+2s]/9 \\
F_2^{\nu N}=F_2^{{\overline{\nu }}N} & = & x[(u+\overline{u})+(d+\overline{d}%
)+2s] \\
x[F_3^{\nu N}+F_3^{{\overline{\nu }}N}]/2 & = & x[u+d-\overline{u}-\overline{%
d}]
\end{array}
\label{DISparton}
\end{equation}

As noted in \cite{MRSA}, these four quantities can be used to extract four
combinations of parton distributions, {\it e.g.,\ } $u+\overline{u}$, $d+%
\overline{d}$, $s$, and $\overline{u}+\overline{d}$, or, equivalently, $u+d$%
. In particular, these four combinations are sufficient for examining the
question of the breaking of SU(3) flavor symmetry of the quark-antiquark
sea. Utilizing the equations given above, the strange sea may be expressed
as
\begin{equation}  \label{strange}
xs=\frac 56F_2^{\nu N}-3F_2^{\mu N}.
\end{equation}
Since the right-hand-side appears as a small difference between two much
larger numbers, the relative uncertainty becomes large and, furthermore, is
sensitive to the overall systematic errors of the experiments---even
though, in recent high precision experiments, the latter have been reduced
to a level sufficient for the application of this relation. A more direct
measure of the strange quark sea is provided by the $\nu $ production of
charm. Unfortunately, data on this process have not yet been made available
in a form independent of experimental corrections. This issue will be
discussed in section \ref{subs:CTEQ1}.

The question of SU(2) breaking in the sea is not directly addressed by the
types of data listed above. Some information is provided by the Gottfried
integral which takes the form
\begin{equation}  \label{GS}
I(a,b)=\int_a^b[F_2^{\mu p}-F_2^{\mu n}]\frac{dx}x.
\end{equation}

The NMC Collaboration has measured \cite{NMC-GS} $I(.004,.8)=0.236\pm 0.008$%
. In lowest order one has
\begin{equation}  \label{GS_PM}
I(0,1)=\frac 13-\frac 23\int_0^1(\overline{d}-\overline{u})dx.
\end{equation}
The experimental result cited above indicates that $\overline{d}>\overline{u}
$ when integrated over $x$. However, information on the $x$ dependence of
this SU(2) breaking must be found from another source.

Lepton-pair production (LPP, or the Drell-Yan process) provides direct
information on the anti-quark distributions as well as the difference
between $u$ and $d$ quarks. For simplicity, consider the cross-section
\begin{equation}
\frac{d\sigma }{dQ^2dy}|_{y=0}  \label{xsec}
\end{equation}
for LPP in proton collisions on an isoscalar target. In lowest order,
retaining only the light quark and antiquark contributions, this cross
section is proportional to the following product of parton distributions:
\begin{equation}
\Sigma _{\mu \mu }=(4u+d)(\overline{u}+\overline{d})+(4\overline{u}+%
\overline{d})(u+d)  \label{sigma1}
\end{equation}
where each of the distributions is evaluated at $x=Q/\sqrt{s}$. Note that
all terms on the right-hand side are directly proportional to anti-quark
distributions (in contrast to DIS where $\bar q(x)$ usually is submerged
under $q(x)$ for a large part of the x-range). Eq.(\ref{sigma1}) can be
rewritten as
\begin{equation}
\Sigma _{\mu \mu }=5(\overline{u}+\overline{d})(u+d)+\frac 32(\overline{u}+%
\overline{d})[(u+\overline{u})-(d+\overline{d})]+\frac 32(\overline{u}-%
\overline{d})[(u+d)-(\overline{u}+\overline{d})].  \label{sigma2}
\end{equation}
In principle, all of the terms except $(\overline{u}-\overline{d})$ are
constrained by the deep inelastic data. Therefore, the lepton pair data
provide a direct measure of the SU(2) breaking of the sea, $i.e.\ (\overline{%
u}-\overline{d})$, when used in conjunction with the deep inelastic data. In
fact, the E-605 data on $d^2\sigma /dQ^2dy$ used in this analysis cover a
range in $y$ (centered about zero). This provides even more information than
the $y=0$ case shown above -- since the y-dependence extends the $x$ range
through the relation $x_{1,2}=\sqrt{Q^2/s}e^{\pm y}$ -- but the principle is
the same.

All the CTEQ analyses result in substantial SU(2) breaking due to the use of
the full range of DIS and LPP data. Since the E-605 data constrain the PDF's
over a range in $x$ covering approximately 0.10 - 0.6 (when the $y$-range is
taken into account), the SU(2) breaking effects observed are reliable only
over this range. To extend these results to lower values of $x,$ additional
experimental measurements will be needed.

Recently, the NA51 experiment\cite{NA51} measured the asymmetry between the
cross section for producing lepton pairs from proton and neutron targets,
designed to probe directly the quantity $(\overline{u}-\overline{d})$. As
shown in \cite{ElSt}, this quantity can be written as
\begin{equation}  \label{ADY}
A_{DY}=\frac{(4u_v-d_v)(\overline{u}-\overline{d})+(u_v-d_v)(4\overline{u}-%
\overline{d})}{(4u_v+d_v)(\overline{u}+\overline{d})+(u_v+d_v)(4\overline{u}+%
\overline{d})}
\end{equation}
where the subscript $v$ denotes a valence distribution. The NA51 result is $%
A_{DY}=-0.09\pm 0.028$ at $y=0$ and $Q/\sqrt{s}=0.18$, where the statistical
and systematic errors have been added in quadrature. Comparison with Eq.(\ref
{ADY}) shows that since $u_v/d_v\approx 2$, one must have $\overline{u}<%
\overline{d}$. This is consistent with the sign of the breaking indicated by
the Gottfried sum result.

Also of interest is the lepton charge asymmetry recently observed in $W$
production by the CDF Collaboration. Consider the charge asymmetry of $W$
production (before decaying into leptons), defined as
\begin{equation}  \label{Wasym}
A_W(y)=\frac{d\sigma ^{+}/dy-d\sigma ^{-}/dy}{d\sigma ^{+}/dy+d\sigma ^{-}/dy%
}
\end{equation}
where the superscript denotes the charge of the $W$. For $\overline{p}p$
collisions in leading order parton model, $A_W(y)$ is given approximately by
\begin{equation}  \label{Wasym2}
A_W(y)\approx \frac{u(x_1)d(x_2)-d(x_1)u(x_2)}{u(x_1)d(x_2)+d(x_1)u(x_2)}
\end{equation}
where $x_{1,2}=x_0e^{\pm y}$ and $x_0=M_w/\sqrt{s}$. Letting $R_{du}=d/u$,
one can write
\begin{equation}  \label{Wasym3}
A_W(y)=\frac{R_{du}(x_2)-R_{du}(x_1)}{R_{du}(x_2)+R_{du}(x_1)}.
\end{equation}
As noted in Ref.\cite{ElSt}, in the region of small $y$ (where $%
R_{du}(x_1)\approx R_{du}(x_2)\approx R_{du}(x_0)$) this asymmetry is
directly proportional to the {\em slope of the ratio} $R_{du}$ in $x$:
\begin{equation}  \label{Wasym4}
A_W(y)\approx -x_0y\frac{dR_{du}}{dx}(x_0)/R_{du}(x_0).
\end{equation}
For the CDF experiment, $x_0=0.044$ and $|y|<2$, thereby providing
information on the ratio of the $d$ and $u$ distributions in the region of $%
x $ of ($0.01,0.2$). Actual data on this process are for the corresponding
decaying lepton asymmetry, so the above discussion is relevant only on the
qualitative level since Eqs.(\ref{Wasym2}-\ref{Wasym4}) are considerably
smeared when applied to the measured leptons.

As mentioned earlier, in addition to these simple parton model relations,
some observables can be sensitive to parton distributions through NLO
effects. Two examples come readily to mind: the precise data on DIS place
important constraints on the gluon distribution $g(x,Q)$ in the region $x<0.2$
(not covered by current fixed-target direct-photon data) through the $Q$%
-dependence of the structure functions; and LPP data provide additional
constraints on $g(x,Q)$ through the ``Compton-scattering'' mechanism. These
examples caution us against taking simple parton relations too literally
under all circumstances.

\subsection{Choice of Parametrization \label{subs:parametrization}}

We now address the issue of the parametrization of the initial PDF's at some
$Q_0$ which serves as the non-perturbative input to the global analysis. The
forms chosen must be flexible enough to account for all experimental input,
if possible, yet they should not be under-constrained. Considering the
current status of the experimental evidence as discussed above, the
parametrization must allow for breaking of both SU(3) and SU(2) flavor
symmetry. Our input parton distributions are parametrized at $Q_0=1.6{\ {\rm %
GeV}}$ (which coincides which the charm threshold we use, see below). The $Q$%
-dependence of the parton distributions is generated by QCD-evolution using
two-loop expressions for the splitting functions and running coupling. In
general, the \mbox{$\overline{\rm MS}$} factorization scheme is used
although, in response to the need for DIS-scheme and leading-order (LO)
calculations, we also generate equivalent parton distributions in these
schemes. The heavy quark thresholds are taken as 1.6 and 5.0 GeV for the $c$
and $b$ quarks, respectively, and the heavy quark distributions are
generated using massless evolution starting from a boundary condition of a
vanishing PDF at the appropriate threshold equal to the corresponding quark
mass. The renormalization scheme on which this definition of heavy quark
parton distribution functions is based has been formulated precisely in Refs.%
\cite{ColTun87,ACOT94}. In principle, it is possible to have non-zero
heavy-quark distributions at threshold -- {\it e.g.} to have some
``intrinsic charm'', as has been suggested occasionally in the literature.
We do not include this possibility for lack of positive experimental
evidence at this time.

The functional forms used for the initial parton distributions in the three
rounds of CTEQ analyses vary slightly. We give here the explicit expressions
used in the most current CTEQ3 analysis:
\begin{equation}
\begin{array}{rcl}
xu_v & = & a_0^ux^{a_1^u}(1-x)^{a_2^u}(1+a_3^ux^{a_4^u}) \\
xd_v & = & a_0^dx^{a_1^d}(1-x)^{a_2^d}(1+a_3^dx^{a_4^d}) \\
xg & = & a_0^gx^{a_1^g}(1-x)^{a_2^g}(1+a_3^gx) \\
x(\overline{d}+\overline{u})/2 & = &
a_0^{+}x^{a_1^{+}}(1-x)^{a_2^{+}}(1+a_3^{+}x) \\
x(\overline{d}-\overline{u}) & = &
a_0^{-}x^{a_1^{-}}(1-x)^{a_2^{-}}(1+a_3^{-}x) \\
xs & = & \kappa \cdot x(\overline{d}+\overline{u})/2
\end{array}
\label{param}
\end{equation}
The coefficients $a_0^u$ and $a_0^d$ are fixed by the number sum rules for
the valence quarks\footnote{%
For our choice of functional form, Eq.~\ref{param}, $a_0^u$ and $a_0^d$ can
be expressed as combinations of Euler Beta functions, {\it e.g.} $%
a_0^u=2/[B(a_1^u,a_2^u+1)+a_3^uB(a_1^u,a_2^u+a_4^u+1)]$.} and the gluon
normalization coefficient $a_0^g$ is fixed by momentum conservation.
Furthermore, with the data currently available it is not possible to
separately determine the low-$x$ behavior for the sea and gluon
distributions, so we have chosen $a_1^g=a_1^{+}$ and set the strange quark
distribution to be proportional to the average non-strange sea. We have also
fixed $\kappa =1/2$ in most of our fits since the resulting $s(x,Q_0)$
agrees well with the recently published NLO strange quark distribution
measured in the most accurate dimuon neutrino experiment \cite{ccfr2mu2}.
(Deviations from these choices used in earlier CTEQ1 and CTEQ2 analyses will
be noted in the next section.) Further reduction of independent parameters
could be achieved by assumptions such as $a_1^u=a_1^d=a_1^{-}$ (motivated by
Regge exchange considerations). The viability of such assumptions needs to
be tested during the process of the global analysis.

In practice, the series of CTEQ analyses adopted the procedure of starting
with a sufficient number of parameters to establish a good fit, then
systematically reducing that number to eliminate extraneous degrees of freedom
while maintaining good agreement with data. In the most recent CTEQ analyses
we found it possible to obtain excellent overall fit using only 15
independently adjustable {\em shape parameters} to describe the input
distributions (see Sec. \ref{subs:CTEQ3} and Table \ref{tbl:param} for
details). In addition, there are individual normalization parameters for
each experiment (constrained by appropriate experimental errors, as
described earlier), the value of $\Lambda _{QCD}$, and the value of the
parameter associated with the theoretical scale uncertainty in direct photon
calculations discussed in Sec.\ref{subs:fitting}.

Applying the PDF's obtained here to generate predictions for processes at
new facilities in regions of $x$ and $Q^2$ beyond those covered in the
current global analyses necessarily entails extrapolations in these
variables. If one is interested in a region of $x$ below that which was
fitted, but at a higher value of $Q^2$, the ``feed down'' property of the
evolution equations (due to the parton splitting process) provides
reasonably reliable extrapolations (cf. \cite{OwTu}) -- provided the input
distribution functions in this $x$ region are relatively smooth (hence the
result is dominated by the nature of the splitting kernel). On the other
hand, if one is interested in small $x$ and moderate $Q^2$, where the PDF's
are still dominated by the input functions, the results are in fact only
extrapolations, not constrained either by theory or experiment. It is thus
important to chose functional forms that smoothly extrapolate into such
regions while simultaneously acknowledging the inherent risk of such
extrapolations. Sometimes, a given functional form can lead to unintended
behavior of the parton distributions beyond the region where data exist.
These considerations must be kept in mind, as the parametrization of the
non-perturbative initial parton distributions, although guided by certain
qualitative ``theoretical considerations''(many of which have had to be
abandoned in recent years in the face of new experimental results), is
ultimately dictated by data and by experience gained in previous global
analyses.

The choices shown above are certainly not unique and do, in fact, differ
slightly from those used in other work, both by us and by other groups \cite
{MRSA}. It is possible to generate fits of comparable quality (in the sense
of least-$\chi ^2$) using somewhat different functional forms as long as
both forms can parametrize the requisite parton distribution shapes to
account for current data. In that case, any remaining difference in the
parton distributions can only be resolved by future experiments.\footnote{%
A detailed study of this issue under current experimental conditions will be
reported in a separate paper.}



\section{Results on Parton Distributions\label{sec:PDFs}}

Three rounds of global analysis based on the general procedures described
above have been completed by the CTEQ collaboration. A short report on the
CTEQ1 analysis has already been published \cite{CTEQ1}. Aside from obtaining
several up-to-date sets of parton distributions (the ``CTEQ1
distributions''), this analysis uncovered an unexpected inconsistency among
existing experiments concerning the flavor dependence of the sea quark
distributions. We briefly discuss the relevant points and subsequent
developments on this issue in the next subsection. The advent of new data
from HERA along with an alternative treatment of the strange sea led to the
development of the CTEQ2 distributions which were made available in the Fall
of 1993. These distributions are described in Sec. \ref{subs:CTEQ2}. Recent
lepton pair asymmetry data from NA51 and W-decay lepton asymmetry data from
CDF prompted refinements of the analysis, resulting in a new set of CTEQ3
distributions which we discuss in detail in Sec. \ref{subs:CTEQ3}.\footnote{%
Computer programs for generating all the CTEQ parton distributions described
below are available from H.L. Lai (Lai\_H@msupa.pa.msu.edu) or W.K. Tung
(Tung@msupa.pa.msu.edu) upon request.} Comparisons with other distributions
are presented in Sec.\ref{sec:compare}.

\subsection{CTEQ1 Parton Distributions\label{subs:CTEQ1}}

The CTEQ1 analysis \cite{CTEQ1} was based on data on cross-sections and
structure functions available at the end of 1992. The list of data sets used
is given in Table~\ref{tbl:dataset} with ``1'' marked in the final column.
Very good fits to this wide range of data were obtained---both the overall $%
\chi ^2$ and the $\chi ^2$ distribution among the experimental data sets
indicate a remarkable degree of consistency and are in much better
quantitative agreement with the available data than previous global fits.
Five sets of parton distributions representing two best fits in the $%
\overline{MS}${\rm \ }and DIS scheme (CTEQ1M and CTEQ1D), one fit with a
``singular'' gluon distribution (CTEQ1MS), one with $\Lambda _{QCD}$ fixed
at a higher (``LEP'') value (CTEQ1ML), and one suitable for leading order
calculations (CTEQ1L) were obtained. See Ref.~\cite{CTEQ1} for details.

One disturbing feature of the CTEQ1 parton distributions was that the
strange quark distribution $s(x,Q)$ obtained was considerably larger in the $%
x<0.1$ region then those obtained from leading-order parton model analysis
of the neutrino dimuon production data \cite{cdhs2mu,charm2mu,ccfr2mu1}. It
was pointed out that this $s(x,Q)$ behavior follows necessarily from the
high precision input data sets on total inclusive structure functions
measured by the CCFR and NMC collaborations through the familiar (``charge
ratio'') parton model identity $\frac 56F_2^{\nu N}-3F_2^{\mu N}=s(x,Q)+\,$%
{\em small corrections}, cf. Eq.(\ref{strange}). As remarked earlier,
although this combination of structure functions entails using the (small)
difference between two larger numbers, the quoted experimental statistical
and systematic errors of the relevant high precision DIS experiments are
even smaller, hence enabling this relation (which is implicitly embedded in
the global analysis calculations) to play a decisive role in the
determination of $s(x,Q)$.

The apparent disagreement with the dimuon results on $s(x,Q)$ imply either
the theoretical input (to the global analysis or to the dimuon analysis) has
deficiencies, or some experimental data sets are inconsistent with each
other within the quoted errors. Although our global analysis, by itself,
cannot resolve this dilemma, it was the insistence on taking available data
and their quoted errors seriously which resulted in uncovering this
controversial issue.\footnote{%
To avoid this inconsistency, one has to either arbitrarily enlarge the
quoted experimental errors or overlook (and accept) statistically
significant inconsistent fits.} Ref.~\cite{CTEQ1} suggested careful
examination of all possible theoretical and experimental sources of this
disagreement. Subsequently, CCFR has reanalyzed their dimuon data \cite
{ccfr2mu2} using the NLO formalism of \cite{AOT90,ACOT94} (which is more
consistent with our theoretical framework), resulting in a modified strange
quark distribution. Nonetheless, the above disagreement persists.

On the theory side, the treatment of heavy quark production channels in the
total inclusive structure functions $F_{2,3}^{\nu N}$ and $F_2^{\mu N}$ in
all existing global analysis work is done using the familiar zero-mass
formalism plus a leading-order ``slow-rescaling'' correction prescription
---hence is not truly consistent with the overall NLO and dimuon analyses. A
proper method to treat this problem now exists, cf. Refs.\cite{AOT90,ACOT94}%
. The implementation of this improved theoretical calculation is underway by
the CTEQ group.

On the experimental front, there is considerable sentiment that information
obtained on $s(x,Q)$ from neutrino dimuon data should be more reliable than
that from the difference of $F_2^{\nu N}$ and $F_2^{\mu N}$ obtained in
total inclusive measurements ---in spite of the quoted errors. If this is
the case, then there exists some inconsistency in currently available data
on $F_2^{\nu N}$ and $F_2^{\mu N}$ in the $0.01<x<0.1$ region \cite{Shaevitz}%
,\cite{Voss}. At least one of these data sets needs to be reassessed,
particularly concerning systematic errors.

The neutrino dimuon results were not included in the CTEQ1 analysis because
experimental data in this process are not, so far, available in the form of
detector-independent physical quantities (i.e. structure functions) which
can be included in a global analysis treating all data on the same footing.
In view of the resulting inconsistency, the CTEQ2 analysis takes the
complementary approach of making direct use of the strange quark
distribution function obtained by the CCFR collaboration from their parton
model analysis of the dimuon data, thereby setting this process apart from
all the other experimental input. Obviously, neither approach is completely
satisfactory. Eventually, we need to understand the source of the
inconsistency, and perform a consistent global analysis including measured
dimuon structure functions, thereby avoiding a separate treatment of the
strange quark.

There is another process which is potentially sensitive to the size of the
strange sea. $W-$boson plus charm associated production at hadron colliders
involves a term which is directly proportional to the strange quark sea.
Estimates for this process show that it may be possible to provide some
limits on the strange/non-strange ratio as further data are accumulated \cite
{W-charm}. In addition, a next-to-leading-order calculation of this process
is in progress \cite{W-charm-nlo}.

\subsection{CTEQ2 Parton Distributions\label{subs:CTEQ2}}

The CTEQ2 analysis was initiated after the first measurement of $%
F_2^{ep}(x,Q)$ from HERA became available \cite{HERA93}. These new data not
only extended the measured range of $x$ by two orders of magnitude; they
also offered the possibility of formulating the global analysis in an
alternative way in the face of the dilemma exposed by the CTEQ1 study. The
HERA data provide very useful constraints on the small-$x$ behavior of the
parton distributions even with their relatively large initial errors because
of the extended reach down to $x\sim 10^{-4}$. We therefore modified the
input used in the CTEQ1 analysis by adding the new HERA data in conjunction
with: (i) using a parametrized function $s(x,Q_0)$ obtained by the CCFR
collaboration in NLO QCD analysis which was allowed to vary within an error
band provided by the experiment \cite{ccfr2mu2}; (ii) removing the
conflicting CCFR and NMC $F_2$ data between $x=.01$ and $x=.09$ which forced
the large strange sea through the charge ratio relation, Eq.(\ref{strange}),
in the previous analysis; (iii) including the same fixed-target lepton-pair
and direct photon production data sets; and (iv) adding the new collider
data on lepton-pair production obtained by CDF\cite{CDF-lpp}. The full list
of experiments appears in Table \ref{tbl:dataset} with the last column
marked either 1 or 2.

We obtained global fits to the experimental data mentioned above, again,
with remarkable consistency over all data sets. (See Table \ref{tbl:chisq}
for detailed information on $\chi ^2$ distributions.) Six representative
sets of parton distributions were selected for use in applications.
Following the general CTEQ convention, they are designated as CTEQ2M,
CTEQ2MS, CTEQ2MF, CTEQ2ML (for $\overline{MS}$ best fit, Singular, Flat, and
{\em LEP-}$\Lambda $ respectively)\footnote{%
To be specific: CTEQ2MS (CTEQ2MF) assumes a {\em singular (flat) } small-$x$
behavior of the form $xf(x,Q_0)\sim x^{-0.5}\;(x^0)$ for the sea quarks and
gluons; and CTEQ2ML fixes $\Lambda _5$ at $220$ MeV. For comparison, the
{\em standard} CTEQ2M has $xf(x,Q_0)\sim x^{-0.26}$ and $\Lambda _5=139$ MeV.%
}, CTEQ2L (Leading order best fit), and CTEQ2D (DIS scheme best fit). The
parameters for the initial distribution functions are given in Table \ref
{tbl:param2}.

In comparison to recent experimental data not included in the fit, the CTEQ2
prediction for the charge asymmetry in lepton-pair production $A_{DY}$,
cf. Eq.(%
\ref{ADY}), is small and negative ---in qualitative agreement with the new
NA51 data.\cite{NA51} This is shown in Fig.\ref{fig:NA51}.\footnote{%
The other curves in this figure are obtained from the new CTEQ3M
distributions (to be described in the next subsection) and from the two
recent generations of MRS distributions. Comparisons of these will be
discussed later.\label{fn:seelater}} 
As discussed in Sec. \ref{sec:Relations}, our use of the full set of double
differential cross-section $d^2\sigma /dQ^2dy$ measured by the E605
experiment already constrained the $\bar d-\bar u$ distribution in the $%
0.1<x<0.5$ region. Thus, the (slightly over 1 $\sigma $) agreement of the
CTEQ2 result with the new NA51 data point can be regarded as a reasonable
consistency check. (Other work on parton distributions tend to use the less
comprehensive single differential LPP cross-section $d\sigma /dQ^2$ as a
constraint on fits which include only DIS data, hence do not take advantage
of the full power of the complete E605 data set.)

On the other hand, the recently measured lepton asymmetry in $W$-production $%
A_W(y)$, cf. Eq.(\ref{Wasym}), by CDF conveyed a different message. It was
observed that the predictions of the CTEQ2 distributions were consistently
higher than the data, as shown in Fig.\ref{fig:Wasym}. 
(cf. footnote \ref{fn:seelater}.) Since $A_W(y)$ depends on the $x$%
-variation of the ratio $d/u,$ as discussed in Sec.\ref{sec:Relations}, one
naturally turns to data on the ratio of $F_2^p$/$F_2^n$ in DIS (which also
depends on $d/u$) for a consistency check. It turns out that the CTEQ2
distributions provide an excellent description of the full set of high
precision NMC data on $F_2^p$/$F_2^n.$ In fact, a careful study of the
quality of fits to all experimental data sets (cf. Table \ref{tbl:chisq}) of
the CTEQ2 distributions compared to that of other contemporary distributions
reveals that CTEQ2 gives a much better overall fit (at least in terms of a
substantially lower $\chi ^2)$\footnote{%
To be specific: using our treatment of experimental errors (close to those
specified by the experiments in all cases), the difference in $\chi ^2$ is
of the order 80-90 (for 920 points) which are evenly distributed in one of
the high precision DIS experiments, either BCDMS or CCFR .} even if others
may agree with the specific $A_W(y)$ measurement better. This underlines the
fact that $A_W(y)$ is particularly sensitive to {\em one aspect} of the
PDF's -- the {\em slope} of $d/u$ (cf. Sec.\ref{sec:Relations}) -- which is
not probed by the other experiments. To study the implication of this fact,
we should ask then: Is it possible to vary the CTEQ2 distributions to fit
the $A_W(y)$ data and, at the same time, maintain the same quality of
agreement with all the other experiments? Or, can we reconcile and
understand the interplay of all experiments which play a role in flavor
differentiation of the $u$ and $d$ quarks ---$F_2^p/F_2^n$, E605, $A_W(y)$
and NA51? This question will be addressed in the next section on CTEQ3
analysis.

One may note that the results in Table \ref{tbl:chisq} reveal that the
overall $\chi ^2$ value in the global fit (including the new data sets
mentioned above) for CTEQ2M remains the {\em lowest} even compared to the
two more recent fits which are designed to give better description of the
new data. This fact serves as a reminder that total $\chi ^2$ is not
necessarily the best or only measure of a ``good fit'' in a global analysis.
The balanced distribution of $\chi ^2$'s among data sets, particularly those
which are sensitive to specific features -- such as the $A_W(y)$ measurement
to the $d/u$ ratio (relevant for SU(2) flavor differentiation) -- must also
be taken into account. The new CTEQ3 distributions give a more balanced fit
in this sense at the expense of marginally higher total $\chi ^2$; hence, they
represent an improved general purpose parton distribution set.

Since the CTEQ2 distributions do give such a good global fit to the full
data set, the fine-tuning which leads to CTEQ3 only entails very small
shifts in the $u$ and $d$ quark distributions, as will be shown in the next
two sections. Consequently, for the vast majority of applications which are
not sensitive to the precise distinction between $u$ and $d$ quarks, there
will hardly be any observable differences in practice. In particular, the
special CTEQ2 distributions designed to test specific assumptions, such as
CTEQ2MF (flat) and CTEQ2MS (singular) to map out a range of small-$x$
behavior which bracket the HERA data (cf. Fig.\ref{fig:smallx}) and CTEQ2ML
(large-lambda) which has a higher value of $\Lambda $ with a somewhat
different gluon distribution, remain perfectly valid for their original
purposes.



\subsection{CTEQ3 Analysis and Distributions \label{subs:CTEQ3}}

Previous global analyses have been dominated by experimental data collected
at fixed-target energies. The observed sensitivity of the new CDF data on $%
A_W(y)$ to details of the parton distributions, particularly $u$ and $d$
quarks, ushers in a new stage of global analysis marked by an increasing
role for quantitative measurements at hadron colliders.\footnote{%
Other measurements which will soon play an important role, especially for
probing the gluons, are precise data on direct photon production (including
photon plus jet) and jet cross-sections (including di-jets).} In addition,
with the increased number of physical processes included in the analysis, we
are approaching the point where all parton flavors will be sufficiently
constrained to lead to either an (almost) unique set of PDF's (in the $x$
range covered by the experiments) or evidence for potential inconsistencies.
The detailed CTEQ3 analysis is undertaken to respond to this new development
and to address the related issues discussed at the end of the last
subsection. All data sets listed in Table \ref{tbl:dataset}, including the
recent NA51 and CDF $A_W(y)$ measurements and the final 1994 ZEUS data on $%
F_2(x,Q)$\cite{Zeus94a}, are included in the global fit.

The specific parametrizations for the initial parton distributions (at $%
Q_0=m_c=1.6\,{\rm GeV})$ used in this analysis are discussed in Sec.\ref
{subs:parametrization}. The effect of various choices of functional forms
and the number of independent parton shape parameters on the predicted
behavior of the various processes and on the global analysis have been
extensively studied. We found that: the new data do help constrain the
flavor dependence of the quark distributions, in particular the $u$ and $d,$
much better then before. From these studies, we have chosen a representative
set of new parton distributions -- the CTEQ3 distributions, which give a
best balanced fit to all available data. Details will be described below. In
the next section, we compare these PDF's with other available sets and with
representative experimental data sets. Unresolved issues and assessment on
uncertainties of the PDF's which emerge from this round of detailed analysis
will be discussed in the Sec.\ref{sec:uncertainty}.

Following the general CTEQ convention, the new parton distribution sets in
the commonly used factorization schemes will be referred to as CTEQ3M (%
\mbox{$\overline{\rm MS}$}), CTEQ3D (DIS), and CTEQ3L (Leading-order)
respectively. These three sets are obtained from independent fits to the
same data sets under the same assumptions except the scheme for calculating
the evolution kernel and the hard cross-sections. Thus, they are {\em %
functionally equivalent} in the sense that (when applied in the appropriate
scheme) they yield the same physical cross-sections, within errors, for the
data included in the analysis; they are, however, not {\em algebraically
equivalent} in the sense that they could be obtained from each other by
applying the applicable perturbative transformation formula between the
schemes. The latter is known to be unreliable in many situations where
nominal NLO terms (e.g. those involving a large gluon contribution) are of
comparable numerical size as the LO term (e.g. involving small sea-quarks).
In the ensuing discussions, we shall only mention the CTEQ3M distributions
explicitly.


The parton distribution shape parameters at $Q_0=1.6$ GeV for the CTEQ3
distributions obtained from the global fit are listed in Table \ref
{tbl:param}. During the process of this analysis, we started from the full
set of (18) parameters introduced in Sec.\ref{subs:parametrization}, then
tried to systematically reduce the number of independent parameters while
maintaining the quality of the fit as established by benchmarks from the
starting fits$.$ The final fit involves 15 parton shape parameters, which is
considerably lower than the previous CTEQ analyses, and also lower then the
current MRSA one.

The total $\chi ^2$ is 839 for 850 degrees of freedom, using the data sets
listed in Table \ref{tbl:dataset}. The distribution of the $\chi ^2$ values
among the various processes and data sets is balanced, as summarized in
Table \ref{tbl:chisq} which also show the corresponding $\chi ^2$ values
obtained for some other parton distribution sets (representing both the
current and the previous generation of PDF's) in order to indicate where the
differences between the various sets lie, as already mentioned in the last
subsection. The normalization factors for the various experiments emerging
from the CTEQ3 global fit are given in Table \ref{tbl:ExpNorm}.

To illustrate the quality of the fit, we present in Fig.\ref{fig:muF2p} the
comparison with BCDMS and NMC data on muon $F_2^p(x,Q)$\footnote{%
As mentioned in Secs. \ref{subs:CTEQ1} and \ref{subs:CTEQ2}, data points
from CCFR and NMC structure functions (but not the ratio $F_2^p/F_2^n$) have
been excluded in the analysis, hence are not shown in these plots.
Comparison of the excluded data points with the resulting fit will be
discussed in Sec. \ref{sec:uncertainty}, cf. Fig. \ref{fig:FtoMuNu}. \label
{fn:MissingData}}; in Fig.\ref{fig:F2noF2p} the NMC data on $F_2^p/F_2^n$;
in Figs.\ref{fig:ccfrF2} \& \ref{fig:ccfrF3} the CCFR data on neutrino $F_2$
and $F_3$ (cf. footnote \ref{fn:MissingData}); in Fig.\ref{fig:zeus94} the
latest ZEUS data on $F_2^p$; in Fig.\ref{fig:e605} the double differential
lepton-pair data of E605; in Fig.\ref{fig:dirPh} the combined direct photon
data at $y\sim 0$ from E706, UA6, and WA70. The various data sets appearing on
the same plot in all these figures have been multiplied by offset factors to
avoid overlap; hence the vertical scales are in arbitrary units and they are
not labelled. The ``goodness of fit'' represented by the $\chi^2$ table is
made explicit by these plots. Comparison to the NA51 data
point on lepton-pair charge asymmetry was shown earlier in Fig.\ref{fig:NA51}%
, Sec.\ref{subs:CTEQ2}; and comparison to the CDF W-lepton asymmetry data
was shown in Fig.\ref{fig:Wasym} in Sec. \ref{subs:CTEQ2}.


An overview of the various flavors of CTEQ3M parton distributions at the
scale $Q=5$ GeV is displayed in Fig.\ref{fig:cteq3pdf}. Included (near the
bottom of the figure) is the difference between $\bar d$ and $\bar u$
distributions, which has been a subject of much attention in the last few
years. The fact that we can now investigate quantitatively the behavior of
such a small difference illustrates the significant progress made possible
by recent high precision experiments and accurate calculations. We will
discuss the uncertainty on this difference later.

Concerning the global analysis which lead to the CTEQ3 parton distributions,
we notice that:

\noindent \vspace{0.5ex} \ $\circ $\ The value of $\Lambda _{QCD}^{5\ fl}$
---158 MeV, obtained in this round of analysis (cf. Table \ref{tbl:param}),
is similar to the values obtained in previous global fits. It corresponds to
a value of $\alpha _s(M_Z^2)$=0.112, in agreement with the value determined
from $Q$-dependence of DIS structure functions, but lower than that from
global analysis of LEP data \cite{LEP}, reflecting again the dominance of
DIS in the current global analysis. However, since the value of $\Lambda .$
is correlated with other parameters in the global fit, particularly those
associated with gluon shape which may not be well-determined yet, there is
still a range of uncertainty on $\Lambda $. We found that alternative ways
of treating certain processes, e.g. using a particular fixed scale in direct
photon calculations, can cause $\Lambda $ to shift (usually to higher
values) by 30-40 MeV.

\noindent \vspace{0.5ex} \ $\circ $\ The reach into the small-$x$ ($%
10^{-2}-10^{-4}$) range, and the recent reduction in errors, provided by the
HERA experiments put rather stringent constraints on the effective power $%
a_1\ (-0.35<a_1<-0.25)$ for the sea quarks, cf. Eq. (\ref{param}). The need
to vary this parameter over a certain assumed range, as done in the past,
has diminished. To show the progression of development, Fig.\ref{fig:smallx}
plots the recent ZEUS data on $F_2$ as a function of $x$ at $Q^2=15$ GeV$^2$
compared to CTEQ3M and some previous distribution sets which assume $a_1=0$
(MRSD0', CTEQ2MF) or $a_1=-0.5$ (MRSD-', CTEQ2MS), and which either came
before the advent of any HERA data (MRSD) or were constrained by the early
HERA data (CTEQ2). We see that the MRSD distributions are now away from
current data; whereas the two CTEQ2 sets now bracket the new data points
(rather than ``fit'' them).


It is important to bear in mind that the values quoted for $a_1$ from our
fit, as for others, is applicable at the specified scale $Q_0$ only$\ (1.6$
GeV for CTEQ3). The evolution of the parton distributions with increasing
scale to an ever softer (i.e. singular) shape will cause this effective
power to increase in absolute value. Thus, comparison with ``theoretical
expectations'' of small-x behavior for fixed (but unspecified) $Q,$ such as
those from the BFKL hard pomeron \cite{BFKL}, is inherently of limited
validity.

\noindent \vspace{0.5ex} \ $\circ $\ A new feature of the CTEQ3 (and CTEQ2)
analysis is the inclusion of a theoretical parameter representing the
uncertainty associated with the choice of scale in direct photon
calculations (cf. Sec.\ref{subs:fitting}). The best estimate of this
parameter which gives the optimal overall fit is in the range $\mu
/p_T=(0.4-0.5)$ which is quite reasonable.

Unlike in the past, where within the \mbox{$\overline{\rm MS}$} scheme
some alternative sets reflecting certain uncertainties\footnote{%
Such as small-$x$ behavior and the value of $\Lambda _{QCD}$ (e.g. CTEQ2MS,
CTEQ2MF, and CTEQ2ML).} were also given, we have restricted the CTEQ3
distributions to the three equivalent sets (3M, 3D and 3L) mentioned above
since: (i) these uncertainties are steadily decreasing as progress is been
made; and (ii) as discussed in the last subsection, for making comparative
studies, the alternative CTEQ2 parton sets (2MF, 2MS \& 2ML) still serve the
original purposes quite adequately, as the transition to the new version
only entails certain fine-tuning which does not affect those purposes (e.g.
see Fig. \ref{fig:smallx} and the discussion on small-$x$ behavior above).

\section{Comparisons of Parton Distributions and Recent measurements\label
{sec:compare}}

To see the status of global QCD analysis and the recent progress, we compare
the CTEQ3M parton distributions to the current MRSA set and to the earlier
CTEQ2M and MRSD-' sets. We limit the comparison to these parton distribution
sets since they have been determined in a program comparable in scope to
that which has been described here.

Figs.\ref{fig:uvl}-\ref{fig:str} display the $u_v(x,Q),\ d_v(x,Q),\ g(x,Q),\
\bar u(x,Q),\ \bar d(x,Q),\ s(x,Q)$ distributions respectively from the four
sets of PDF's at $Q=5$ GeV in the range $10^{-4}<x<0.8.$ We observe that:

\noindent \vspace{0.5ex} \ $\circ $\ The small spread between the previous
generation CTEQ2M and MRSD-' valence quarks (curves with dots in Figs.\ref
{fig:uvl} \& \ref{fig:dvl}) has been noticeably narrowed in the current
round of analyses given by CTEQ3M (solid) and MRSA (dashed). The $u_v(x,Q)$
and $d_v(x,Q)$ distributions are now very well determined indeed throughout
the range where they are not vanishingly small. 

\noindent \vspace{0.5ex} \ $\circ $\ As shown in Fig.\ref{fig:glu}, the
gluon distributions from the three sets incorporating HERA data in the fit
-- CTEQ2M, CTEQ3M and MRSA -- are also in close agreement. The more singular
behavior of MRSD-' is due to the input condition without the benefit of
data. We will discuss the uncertainty on the gluon distribution later in
Sec. \ref{sec:uncertainty}. 

\noindent \vspace{0.5ex} \ $\circ $\ For the $\bar u(x,Q)$ and $\ \bar
d(x,Q)\ $distributions, shown in Fig.\ref{fig:ubr} and Fig.\ref{fig:dbr},
the MRSA distributions are somewhat higher than the CTEQ ones, even though
they are both determined mostly by the same HERA data. The reason lies
mainly with the different normalization factors used by the different fits.
(Cf. Table \ref{tbl:ExpNorm}.) This difference arises from the different ways
the two groups treat experimental uncertainties, especially the normalization,
in their respective fits (cf. detailed discussion in Sec.\ref{subs:fitting});
and it is also influenced by our exclusion of the controversial CCFR and
NMC data points below $x=0.09$. 

\noindent \vspace{0.5ex} \ $\circ $\ The differences in the strange
distribution, shown in Fig.\ref{fig:str}, are entirely due to differences in
input assumptions. The CTEQ2M input distribution is taken from the $s(x,Q)$
distribution furnished by the recent CCFR experiment on neutrino dimuon
production \cite{ccfr2mu2}; the other three used the constraint $%
s(x,Q)=(\bar u(x,Q)+\ \bar d(x,Q))/4$ which is consistent with the above
data, within errors, in the measured range $0.015<x<0.3$. The small-$x$
extrapolations follow the functional forms assumed. 

We now discuss briefly the comparison of recent NA51 and CDF charge
asymmetry data which motivated the new round of analyses with results
obtained from these distributions. The new parton distributions MRSA and
CTEQ3M use these data as part of the input and, hence, their agreement with
data is expected. It only remains to understand the changes these new data
brought about in the parton distributions.

In Fig.\ref{fig:NA51} the result \cite{NA51} for $A_{DY}$ is compared to the
results of different fits. Although the data set only consists of one single
point, it obviously has a major impact on the MRS analyses. The effect on
CTEQ analysis is less dramatic (for reasons discussed in Sec.\ref{subs:CTEQ2}%
), but still substantial. Fig. \ref{fig:Wasym} shows a comparison of the
same fits to the $W$ decay lepton asymmetry data \cite{CDF-W} on $A_W$ . The
impact here is mainly on the CTEQ analyses with the much improved agreement
of the CTEQ3M distributions compared to the CTEQ2M distributions. Both MRS
sets fit this data set well.

Although long-established DIS data on proton and nuclear targets, along with
lepton-pair production data provide the main source of information on the $u$
and $d$ quarks, these two recent experiments played a surprisingly
significant role in pinning down the details of the distinction between the
two lightest quark flavors. As discussed earlier in Sec.\ref{sec:Relations},
$A_{DY}$ is mainly sensitive to the difference $(\overline{d}-\overline{u}),$
whereas $A_W$ is most sensitive to the $x$-dependence of the ratio $d/u$
which includes both valence and sea. Hence, we show in Fig.\ref{fig:dmu1}
and Fig.\ref{fig:dou} the comparison of these combinations of parton
distributions from the four sets of distributions respectively. The plot of $%
(\overline{d}-\overline{u})$ in Fig.\ref{fig:dmu1} is the most dramatic in
demonstrating the change of our knowledge on parton distributions brought
about by these recent experiments. The large movement of MRSD-' curve toward
MRSA is forced by the NA51 data. The change of CTEQ2M curve toward CTEQ3M is
influenced by the adjustments needed to fit the $A_W$ data, mainly in the
region around $x=0.05$.\footnote{%
The reduction of one parameter in the input functional form for $(\overline{d%
}-\overline{u})$ from CTEQ2 to CTEQ3 also has some influence in the change.}
The $d/u$ plot of Fig.\ref{fig:dou} does not display a significant
difference in the four curves. Nonetheless, close examination of the
differences in the slope of these curves in the region $0.02<x<0.2$ does
bear out the expectations discussed in Sec.\ref{sec:Relations}.




\section{Uncertainties and Challenges \label{sec:uncertainty}}

Since the existing experimental and theoretical input to global QCD analyses
are not quite sufficiently extensive and accurate to determine a unique set
of parton distributions, it is useful to have some feeling about the
uncertainties of the PDF sets. The common practice of assigning
uncertainties according to the spread of some chosen subset of currently
available distributions is quite haphazard, as most published sets are
selected out of many possible candidates; and as PDF's obtained by different
groups are not always comparable because they are based on different
assumptions and inputs. A comprehensive program to systematically assess the
uncertainties of PDF's based on error matrix analysis is a desirable goal,
but rather difficult because of the complexity of the global system. It is
certainly not presently available. In this section, we state the outstanding
problems in the determination of parton distributions and describe in
qualitative terms the current uncertainties based on extensive exploratory
work done by the CTEQ group beyond that contained in the three rounds of
specific parton distribution sets bearing the collaboration label. We also
comment on the origin of the observed good agreement as well as some of the
minor differences between the current generation of MRSA and CTEQ3
distributions in order to address the question: to what extent do these
agreements and differences reflect real current uncertainties on the parton
distributions?

\noindent \smallskip\ \ $\circ $\ {\bf The gluon distribution:} It is common
knowledge that the best available handles on the gluon distribution are the $%
Q$-dependence of the DIS structure functions and cross-sections for direct
photon production, although it affects all QCD processes---at least through
evolution of all parton distributions and NLO hard cross-sections. These two
processes complements each other. The DIS data are quite precise; but the
measurement is ``indirect'' (i.e. through QCD evolution only) -- hence it is
applicable only in the smaller $x$-region where the influence of the gluon
distribution on the measured structure functions can be seen. The direct
photon measurement is ``direct'' but, so far, available data still have
large errors and theoretical uncertainties are greater. The ``good
agreement'' between the CTEQ and MRS gluon distributions shown in Fig. \ref
{fig:glu} is not evidence for a well-determined $G(x,Q),$ it merely reflects
the common assumptions made by the two groups. For instance, the agreement
below, say $x\sim 0.05$, can be attributed to the fact that the most
important gluon shape parameter which governs its small-$x$ behavior -- $a_1$
in the factor $x^{a_1}$ (cf. Eq.(\ref{param})) -- is assumed to be the same
as that of all the sea-quarks (which is rather well determined by the new
HERA data) by both. This is only an assumption. To assess uncertainties, we
need to go beyond the standard sets.

For moderately large $x$, $G(x,Q)$ should be determined by the direct photon
data used in the global analysis. Fig.\ref{fig:gluB}a displays the gluon
distributions in the range $0.1<x<0.6$ at $Q=5$ GeV from CTEQ3, MRSA and the
alternative CTEQ2 sets which were designed to explore some aspects of parton
distribution uncertainties. The CTEQ2ML set (with a larger $\Lambda $ value
which is closely correlated to gluon behavior) and the CTEQ2MS \& CTEQ2MF
sets (with a different small-$x$ behavior which affects all $x$-ranges by
the momentum sum rule) give a better indication of the range of possible
gluon behavior. The CTEQ analyses used all available fixed-target photon
data sets -- WA70, E706, and UA6. We see that the range of variations are
fairly large. This is because both point-to-point and overall normalization
errors on all these data are still large, and theoretical uncertainties
(partially taken into account in the CTEQ analyses) are not yet under full
control.

The behavior of $G(x,Q)$ in the small-$x$ region is a wide open problem.
Under the commonly made assumption that the $a_1$ parameter for the gluon is
the same as for the sea quarks, Fig. \ref{fig:gluB}b show the same gluon
distribution sets as above over the extended range $10^{-4}<x<0.6$. Since
CTEQ2MF and CTEQ2MS envelop the current HERA data in the small-$x$ range,
this plot gives a reasonable representation of the uncertainty under the
given assumption. However, the possibility that the gluon distribution may
have a different behavior must be kept in mind unless it is ruled out by
future experiment.

More detailed analysis of direct photon data is needed both to achieve
better accuracy and to resolve a possible theoretical problem with the shape
of the $p_t$ distribution observed in existing experiments.\cite{CTEQphoton}
New collider data from the CDF collaboration covering a much smaller $x$
range are becoming available and better fixed-target measurements are
anticipated from the on-going analysis of the E706 experiment. In addition,
a wealth of data on jet-production from CDF and D0, which are even more
sensitive to gluons, are also becoming available. This promises to be an
active area of investigation to gain information on $G(x,Q)$ in hadron
colliders. These efforts will complement well parallel ones actively pursued
at HERA.

\noindent \smallskip\ \ $\circ $\ ${\bf d(x,Q)/u(x,Q):}$ The $d/u$ ratio not
only directly impacts on the W-charge asymmetry ($y$-dependence), it also
has an influence on the $p_T$ distribution of decaying leptons in
W-production which is critical to the understanding of precision measurement
of the $W$ mass. How can one assess the uncertainties on the parton
distributions which affect QCD predictions on these important quantities? As
mentioned earlier, using the range spanned by some set of canned
distributions which fit the $A_W$ data to varying degrees of goodness as an
estimate on the uncertainty is not a satisfying strategy. In the current
round of CTEQ analysis, we have performed a number of studies to explore
this problem.

As an example, we show in Fig.\ref{fig:W}a two fits to the global data in
addition to CTEQ3M with the requirement that the ``upper'' one gives rise to
a W-lepton asymmetry $A_W$ about one standard deviation above the CDF data,
and the ``lower'' one to values of $A_W$ one standard deviation below the
CDF data. The slope of the $d/u$ ratio from these three sets of parton
distributions are shown in Fig.\ref{fig:W}b. Since all three sets give rise
to comparable fits to the rest of the global data sets (with CTEQ3M being
the best fit), the differences exhibited here perhaps represent more
realistically the uncertainty associated with this quantity.

The W-mass measurement, although also sensitive to the $d$ and $u$
distributions, is not dependent on this same quantity. To arrive at a
meaningful assessment of the uncertainties due to parton distributions, it
is desirable to perform a similar study as above but focused on the $p_t$
spectrum on which the mass determination depends. Such a study is underway.

\noindent \smallskip\ \ $\circ $\ {\bf SU(2) Flavor asymmetry of sea quarks
-- $(\bar d-\bar u)$:} As noted in the previous section, important progress
has been made on the difference between $\bar u$ and $\bar d$ quarks. The
main contributing processes are DIS structure functions on proton, deuterium
and nuclear targets, lepton-pair production, and the recent $A_W$
measurement. Do the differences between the MRSA and CTEQ3M ($\bar d-\bar u$%
) function shown in Fig.\ref{fig:dmu1} reflect the current uncertainty on
this quantity? (This question has important bearing on the validity of the
Gottfried Sum Rule.) A closer look at the $\chi ^{2\prime }s$ for these two
sets shown in Table \ref{tbl:chisq} indicates that over half of the
difference comes from the very precise BCDMS sets. A further examination of
the individual data points reveals that much of the extra $\chi ^2$ occurs
at the small-$x$ end of the BCDMS $D_2$ data set. This is the $x$ range
where the ($\bar d-\bar u$) function shows a difference in Fig.\ref{fig:dmu1}%
. This difference may be partially related to our exclusion of the
conflicting CCFR and NMC $F_2$ data below $x=0.09$. One theoretical
uncertainty in this region concerns the size of shadowing corrections to the
deuterium measurement. We have made independent analyses with and without
deuterium corrections (based on Ref.\cite{DeuCor}) and found that the
differences between the resulting parton distributions were insignificant,
and that the above conclusions were unaffected.

In order to obtain a self-contained estimate of the uncertainty of the ($%
\bar d-\bar u$) function, we have performed a series of fits systematically
varying the $a_1$ parameter of this function at $Q_0$. Fig.\ref{fig:dmu2}
shows a band of curves representing the resulting ($\bar d-\bar u$) at $Q=5$
GeV. The overall $\chi ^2$ of these fits, as well as their distribution
among the various experiments, are very similar -- except that those
vanishing faster toward small-$x$ in general give better fits to the CDF $%
A_W $ data. This band plot gives an indication on the uncertainty of ($\bar
d-\bar u$) under the conditions described above. A more detailed study of
this problem and its implication on the Gottfried sum rule is still
underway. We remark that there is a proposed (and approved) experiment at
Fermilab, E866, which will measure this quantity over the kinematic region
in question. With the recent developments, this measurement acquires even
more significance.

\noindent \smallskip\ \ $\circ $\ {\bf SU(3) Flavor asymmetry of sea quarks
-- }${\bf s(x,Q):}$ As discussed in some detail in Sections \ref{subs:CTEQ1}%
{\bf -}\ref{subs:CTEQ3}, the strange quark distribution is not included in
current global analyses on the same footing as the non-strange quarks. The
MRS and CTEQ3 analyses both adopt the assumption that $s(x,Q)=(\bar d+\bar
u)/4$. This is consistent with the neutrino dimuon data; but causes problems
with the available inclusive $F_2^{\nu N}$ and $F_2^{\mu N}$ data at small-$x
$. To see this problem, we show in Fig.\ref{fig:FtoMuNu} the $F_2^{\nu N}$
and $F_2^{\mu N}$ data in the range $0.01<x<0.1$ and $Q^2>4$ GeV$^2$
compared to CTEQ3M and MRSA curves.\footnote{%
The $F_2^{\nu N}$ points are converted from $F_2^{\nu Fe}$ data by using the
ratio $F_2^{\mu D} /F_2^{\mu A}$, (A = Ca, Fe), measured by NMC. The NMC
ratio is consistent with the only, much less accurate, neutrino shadowing
measurement \cite{nushadow}, and with predictions from PCAC \cite{Bell64}.}
The CTEQ2 and CTEQ3 analyses leave out these data points because their
simultaneous inclusion is inconsistent with the assumption made on the
strange distribution in the fitting process, as revealed in the CTEQ1
analysis. The MRS analyses include these data in the fit, seeking a best
comprise. Thus, the MRSA curves are closer to the data points in Fig.\ref
{fig:FtoMuNu}, but at the expense of higher overall $\chi ^2$ --
particularly on the BCDMS measurements. It appears that, this discrepancy
needs to be understood before we can have complete confidence in our
knowledge on $s(x,Q)$.

\section{Summary and Conclusions \label{sec:summary}}

The sequence of analyses reported here give a realistic view of the manner
in which progress in theory and experiment interact as the characteristics
of the various parton distributions are investigated. The latest version of
CTEQ analysis, CTEQ3, provides an excellent description of a wealth of data
covering an extended range in both $Q^2$ and $x$ compared to what was
available just a few years ago. The precision of the data and the diversity
of physical processes together allow detailed investigations of {\em fine
structures} such as the breaking of flavor symmetry in the sea. Where possible
we indicated remaining sources of uncertainty and suggested what types of
data might help to reduce this in the future. One interesting area concerns
the small-$x$ behavior of the gluon and how to reconcile its behavior there
with observables which are sensitive to the behavior at moderate to large
values of $x$. Certainly future collider measurements of jet and photon
production will play a leading role in such studies. The remaining
uncertainties on quark distributions concern detailed flavor
differentiation, particularly among the sea quarks. New measurements on
vector boson production (W-, Z- and continuum lepton-pair) will be valuable,
as illustrated by the first CDF results on $A_w$ and NA51 on $A_{DY}$; and
clarification of the difference between $F_2^{\nu N}$ and $F_2^{\mu N}$ data
in the range $0.01<x<0.1$ is sorely needed.

Related projects which are underway and will be reported separately cover a
range of topics relevant to the global analysis of PDF's. These include a
comprehensive survey of direct photon measurements spanning the range from
fixed target to collider energies, a detailed examination of issues related
to the choice of parametrizations and the effects on the description of
individual experiments, and a method of estimating errors on predictions due
to the uncertainties associated with the parton distribution determinations.

The results presented here should be considered in the same sense as a
snapshot showing the state of the subject at one instant of time. As new
data and calculations become available, the underlying QCD framework will be
ever more critically tested, and further progress toward a unique
determination of parton distributions will be made.


\section*{Acknowledgement}
We would like to thank our colleagues in the CTEQ Collaboration for many
lively discussions and useful suggestions. We also would like to
note that a global analysis project of this scope necessarily
involves indirect contributions from a wide spectrum of sources worldwide
through constant communications with many
physicists, particularly some key members of experiments
whose data are used in the analyses.
These sources are too numerous to name individually.
Likewise, the programs used to perform these analyses, although mostly
developed by us, do contain elements from other sources and involve some
contribution from our CTEQ colleagues, to all of whom we express our thanks.




\newtheorem{Figure}{Figure}
\def\figWasym
{
\begin{Figure}

  {\rm The CDF W lepton charge asymmetry data compared to NLO QCD
results obtained from previous and
current versions of MRS and CTEQ parton distributions. }
  \label{fig:Wasym}
\end{Figure}
}
\def\figDYasym
{
\begin{Figure}
  {\rm A comparison of the data for $A_{DY}$ from NA51 with NLO QCD
results obtained from previous and
current versions of MRS and CTEQ parton distributions. }
  \label{fig:NA51}
\end{Figure}
}
\def\figmuFp
{
\begin{Figure}
  {\rm Comparison of the CTEQ3 fit with $F_2^{\mu N}$ data of BCDMS and NMC
experiments.
The absolute vertical scale is not labelled since
an offset factor has been applied to the various $x$ bins to avoid overlap.}
  \label{fig:muF2p}
\end{Figure}
}
\def\figFnoFp
{
\begin{Figure}
  {\rm Comparison of the CTEQ3 fit with $F_2^{\mu n}/F_2^{\mu p}$ data of the
NMC experiment.}
  \label{fig:F2noF2p}
\end{Figure}
}
\def\figccfrA
{
\begin{Figure}
  {\rm Comparison of the CTEQ3 fit with $F_2^{\nu N}$ data of the CCFR
experiment.
Data points are converted from the measured $F_2^{\nu Fe}$ by using the
ratio $F_2^{\mu D} /F_2^{\mu Fe}$ measured by NMC.}
  \label{fig:ccfrF2}
\end{Figure}
}
\def\figccfrB
{
\begin{Figure}

  {\rm Comparison of the CTEQ3 fit with $F_3^{\nu N}$ data of the CCFR
experiment.
Data points are converted from the measured $F_3^{\nu Fe}$ as for $F_2^{\nu
N}$.}
  \label{fig:ccfrF3}
\end{Figure}
}
\def\figZeus
{
\begin{Figure}

  {\rm Comparison of the CTEQ3 fit with current $F_2^{\mu p}$ data of the ZEUS
experiment.}
  \label{fig:zeus94}
\end{Figure}
}
\def\figFermiDY
{
\begin{Figure}

  {\rm Comparison of the CTEQ3 fit with the double differential cross-section
data of the
E606 lepton-pair production (Drell-Yan) experiment.}
  \label{fig:e605}
\end{Figure}
}
\def\figdirPh
{
\begin{Figure}

  {\rm Comparison of the CTEQ3 fit with $d \sigma / dp_t$ data of three fixed
target
direct photon production experiments.}
  \label{fig:dirPh}
\end{Figure}
}
\def\figcteqpdf
{
\begin{Figure}

  {\rm An overview of all parton distribution functions at $Q=5$ GeV from the
new CTEQ3M analysis. }
  \label{fig:cteq3pdf}
\end{Figure}
}
\def\figuvl
{
\begin{Figure}

  {\rm Comparison of the valence $u$ quark distribution at $Q=5$ GeV from the
current and previous versions of CTEQ and MRS sets.}
  \label{fig:uvl}
\end{Figure}
}
\def\figdvl
{
\begin{Figure}

  {\rm Comparison of the valence $d$ quark distribution at $Q=5$ GeV from the
current and previous versions of CTEQ and MRS sets.}
  \label{fig:dvl}
\end{Figure}
}
\def\figglu
{
\begin{Figure}

  {\rm Comparison of the gluon distribution at $Q=5$ GeV from the current and
previous versions of CTEQ and MRS sets.}
  \label{fig:glu}
\end{Figure}
}
\def\figubr
{
\begin{Figure}

  {\rm Comparison of the sea quark distribution $\bar{u}$ at $Q=5$ GeV from the
current and previous versions of CTEQ and MRS sets.}
  \label{fig:ubr}
\end{Figure}
}
\def\figdbr
{
\begin{Figure}

  {\rm Comparison of the sea quark distribution $\bar{d}$ at $Q=5$ GeV from the
current and previous versions of CTEQ and MRS sets}
  \label{fig:dbr}
\end{Figure}
}
\def\figstr
{
\begin{Figure}

  {\rm Comparison of the strange quark distribution at $Q=5$ GeV from the
current and previous versions of CTEQ and MRS sets}
  \label{fig:str}
\end{Figure}
}
\def\figdmuA
{
\begin{Figure}

  {\rm A comparison of the results for the $\overline d - \overline u$
difference from various sets of distributions. }
  \label{fig:dmu1}
\end{Figure}
}
\def\figdou
{
\begin{Figure}

  {\rm A comparison of the results for the ratio $d/u$ from various
sets of distributions. }
  \label{fig:dou}
\end{Figure}
}
\def\figsmallx
{
\begin{Figure}

  {\rm Comparison of the current ZEUS small-$x$ data at $Q^2=15$ GeV$^2$ with
various
parton distribution sets with different $x$ exponent values.}
  \label{fig:smallx}
\end{Figure}
}
\def\figgluB
{
\begin{Figure}

  {\rm The gluon distribution at $Q=5$ GeV in the ranges -- (a) $0.1<x<0.6$,
and (b)
$10^{-4}<x<0.6$ -- from CTEQ3M, MRSA and from the
alternative CTEQ2 sets which are designed to explore various aspects of parton
distribution uncertainties.}
  \label{fig:gluB}
\end{Figure}
}
\def\figW
{
\begin{Figure}

  {\rm (a) Three fits bracketing the CDF W lepton-asymmetry data;
(b) The corresponding slopes of the $d/u$ ratio of the three sets of parton
distributions. }
  \label{fig:W}
\end{Figure}
}
\def\figdmuB
{
\begin{Figure}

  {\rm A band of $(\bar d-\bar u)$ from a series of global fits which yield
comparable $\chi^2$'s.}
  \label{fig:dmu2}
\end{Figure}
}
\def\figFtoMuNu
{
\begin{Figure}

  {\rm Comparison of (a) CCFR neutrino and (b) NMC muon measurements of $F_2$
in the $0.015<x<0.09$ region compared to MRSA and CTEQ3M curves. These data
points are excluded from the CTEQ2 and CTEQ3 fits since, taken together,
they conflict with the strange quark distribution adopted in the fit.
}
  \label{fig:FtoMuNu}
\end{Figure}
}

\newpage
\centerline{\Large Figure Captions}

\figDYasym\figWasym\figsmallx
\figmuFp\figFnoFp\figccfrA\figccfrB
\figZeus\figFermiDY\figdirPh
\figcteqpdf\figuvl\figdvl\figglu
\figubr\figdbr
\figstr\figdmuA\figdou
\figgluB
\figW
\figdmuB
\figFtoMuNu

\def\tbldataset
{
\begin{table}[p]
\begin{center}
\begin{tabular}{|c|c|c|c|c|c|}
\hline

Process &Experiment &  Observable & Data Points & $\Delta\sigma $ & Set  \\
\hline
DIS & BCDMS & $F_{2 \ H}^{\mu }$ & 168 & .02 & 1 \\ \hline
& & $F_{2 \ D}^{\mu}$ & 156 & .02 & 1 \\ \hline
& NMC &  $F_{2 \ H}^{\mu}$ & 52 & .02 & 1 \\ \hline
& & $F_{2 \ D}^{\mu}$ & 52 & .02 & 1 \\ \hline
& H1 & $F_{2 \ H}^{\mu}$ & 21 & .04 & 2 \\ \hline
& ZEUS &$F_{2 \ H}^{\mu}$ & 56 & .03 & 2 \\ \hline
& CCFR & $F_{2 \ Fe}^{\nu}$ & 63 & .02 & 1 \\ \hline
& & $x \ F_{3 \ Fe}^{\nu}$ & 63 & .02 & 1 \\ \hline
& NMC &  $F_2^n / F_2^p$ & 89 & - & 1 \\ \hline
Drell-Yan & E605 & $s d\sigma / d \sqrt{\tau} dy$ & 119 & .1 & 1 \\ \hline
& CDF & $s d\sigma / d \sqrt{\tau} dy$ & 8 & .1 & 2 \\ \hline
& NA-51 & $A_{DY}$ & 1 & - & 3 \\ \hline
W-prod. & CDF & Lepton asym. & 9 & - & 3 \\ \hline
Direct $\gamma $ & WA70 & $E d^3\sigma / d^3p$ & 39 & .10 & 1 \\
& & $1.0 \ge y \ge -.75$ & & & \\ \hline
& E706 & $E d^3\sigma / d^3p$ & 8 & .15 & 1 \\
& & $y = 0$ & & & \\ \hline
& UA6  & $E d^3\sigma / d^3p$ & 16 & .10 & 1 \\
& & $y = .3$ & & & \\ \hline
\end{tabular}
\end{center}
\caption{
The data sets used in the CTEQ global analyses.  Data sets marked with 1 in the
final column were used for the CTEQ1 and later fits. Those with a 2 or 3 were
added for the CTEQ2 and CTEQ3 analyses respectively.
The column labelled $\Delta\sigma$ gives
the overall normalization systematic error used in defining the $\chi^2$, as
discussed
in the text.}
\label{tbl:dataset}
\end{table}
}

\def\tblparamtwo
{
\begin{table}[h]
\begin{center}
\begin{tabular}{|c|c||c|c|c|c|c|c|}
\hline
Distribution & Parameter & 2M & 2MS & 2MF & 2ML & 2D & 2L \\
\hline
 $xu_v$  & $a_0^u$ & .269 & .268 & .261 & .266 & .307 & .164 \\
\hline
 &$ a_1^u$ &  .278 &  .276 &  .276 &  .289 &  .254 &  .175 \\
\hline
 &$ a_2^u$ &   3.67 &   3.66 &   3.66 &   3.58 &   3.44 &   3.32 \\
\hline
 &$ a_3^u$ &   29.6 &   29.1 &   29.8 &   30.2 &   25.5 &   44.1 \\
\hline
 &$ a_4^u$ &  .807 &  .801 &  .795 &  .799 &  .917 &  .961 \\
\hline
 $xd_v$  & $a_0^d$ & 1.24 & 1.32 & 1.18 & 1.46 & 1.17 & 1.08 \\
\hline
 &$ a_1^d$ &  .521 &  .538 &  .508 &  .565 &  .511 &  .493 \\
\hline
 &$ a_2^d$ &   3.18 &   3.26 &   3.24 &   3.46 &   3.16 &   3.00 \\
\hline
 &$ a_3^d$ &   -0.85 &   -0.84 &   -0.83 &   -0.59 &   -0.60 &   -1.00 \\
\hline
 &$ a_4^d$ &  1.82 &  1.85 &  2.19 &  2.32 &  2.31 &  2.99 \\
\hline
 $xg$ &$ a_0^g$  & .900 & .197 & 3.05 & .825 & .711 & .521 \\
\hline
 &$ a_1^g$ & -.258 & -.500 &  .000 & -.212 & -.240 & -.259 \\
\hline
 &$ a_2^g$ &   5.19 &   3.82 &   6.53 &   4.55 &   4.84 &   4.61 \\
\hline
 &$ a_3^g$ &    5.13 &    5.81 &    2.64 &   12.0 &    7.43 &   16.3 \\
\hline
 &$ a_4^g$ &  1.12 &  .450 &  2.22 &  1.62 &  .960 &  1.24 \\
\hline
$x(\overline d + \overline u)/2$ & $a_0^+$
 & .0825 & .0130 & .2540 & .1139 & .0947 & .1127 \\
\hline
 &$ a_1^+$ & -.258 & -.500 &  .000 & -.212 & -.240 & -.259 \\
\hline
 &$ a_2^+$ &   8.45 &   7.62 &   9.40 &   9.14 &   8.76 &   8.94 \\
\hline
 &$ a_3^+$ &   12.7 &   38.4 &   13.5 &   15.2 &   14.6 &   17.5 \\
\hline
 &$ a_4^+$ &  1.10 &  0.82 &  1.60 &  1.36 &  1.39 &  1.58 \\
\hline
$x(\overline d - \overline u)$ & $a_0^-$
 & .111 & .105 & .114 & .117 & .121 & .103 \\
\hline
 &$ a_1^-$ &  .012 &  .043 &  .085 &  .031 &  .106 &  .043 \\
\hline
 &$ a_2^-$ &   9.53 &  10.00 &   9.71 &   9.95 &   9.00 &   9.87 \\
\hline
 &$ a_3^-$ &  -14.8 &  -15.5 &  -15.7 &  -15.4 &  -15.7 &  -17.7 \\
\hline
 &$ a_4^-$ & 49.4 & 53.8 & 55.7 & 51.7 & 48.2 & 52.3 \\
\hline
 $xs$ & $a_0^s$ & .156 & .152 & .110 & .155 & .140 & .165 \\
\hline
 &$ a_1^s$ & -0.004 &  0.004 & -0.128 &  0.001 & -0.004 & -0.001 \\
\hline
 &$ a_2^s$ &   6.87 &   6.85 &   6.88 &   6.90 &   6.90 &   6.90 \\
\hline
 &$\Lambda^5 (MeV) $
 & 139 & 135 & 135 & 220 & 155 & 143 \\ \hline
\end{tabular}
\caption{
CTEQ2 input parton distribution function parameters (at $Q_0=1.6$ GeV).
The functional form used in CTEQ2 is
$x f = a^f_0 x^{a^f_1} (1-x)^{a^f_2} (1 + a^f_3 x^{a^f_4} )$,
where $f = u_v, d_v, gluon, (\overline d + \overline u)/2$, $s$; and
$x(\overline d - \overline u) = a^-_0 x^{a^-_1} (1-x)^{a^-_2}
(1 + a^-_3 \protect\sqrt x  + a^-_4 x )$.
}
\label{tbl:param2}
\end{center}
\end{table}
}

\def\tblparamthree
{
\begin{table}[h]
\begin{center}
\begin{tabular}{|c|c||c|c|c|}
\hline
Distribution & Parameter & CTEQ3M & CTEQ3D & CTEQ3L  \\ \hline
 $xu_v$ & $a_0^u$ & 1.37 & 1.36 & 1.29 \\ \hline
 &$ a_1^u$ & .497 & .470 & .452 \\ \hline
 &$ a_2^u$ & 3.74 & 3.51 & 3.51 \\ \hline
 &$ a_3^u$ & 6.25 & 6.19 & 6.85 \\ \hline
 &$ a_4^u$ & .880 & 1.04 & 1.11 \\ \hline
 $xd_v$ & $a_0^d$ & .801 & .837 & .858 \\ \hline
 & $a_1^d$ & .497 & .470 & .452 \\ \hline
 &$ a_2^d$ & 4.19 & 4.22 & 4.20 \\ \hline
 &$ a_3^d$ & 1.69 & 2.58 & 2.54 \\ \hline
 &$ a_4^d$ & .375 & .748 & .947 \\ \hline
 $xg$ &$ a_0^g$ & .738 & .595 & .404 \\ \hline
 &$ a_1^g$ & -.286 & -.332 & -.349 \\ \hline
 &$ a_2^g$ & 5.31 & 5.45 & 5.59 \\ \hline
 &$ a_3^g$ & 7.30 & 11.0 & 18.1 \\ \hline
$x(\overline d + \overline u)/2$ & $a_0^+$ & .0547 & .0330 & .0451 \\ \hline
 &$ a_1^+$ & -.286 & -.332 & -.349 \\ \hline
 &$ a_2^+$ & 8.34 & 8.16 & 7.36 \\ \hline
 &$ a_3^+$ & 17.5 & 23.2 & 14.5 \\ \hline
$x(\overline d - \overline u)$ & $a_0^-$ & .0795 & .0702 & .0566 \\ \hline
 &$ a_1^-$ & .497 & .470 & .452 \\ \hline
 &$ a_2^-$ & 8.34 & 8.16 & 7.36 \\ \hline
 &$ a_3^-$ & 30.0 & 27.1 & 29.9\\ \hline
 $xs$ & $\kappa$ & .5 & .5 & .5 \\ \hline
 &$\Lambda^5 (MeV) $ & 158 & 164 & 132 \\ \hline
\end{tabular}
\caption{
CTEQ3 input parton distribution function parameters (at $Q_0=1.6$ GeV). The
functional
forms are described in Sec.\protect\ref{subs:parametrization}. The number of
independant
parton shape parameters is 15.}
\label{tbl:param}
\end{center}
\end{table}
}

\def\tblchisq
{
\begin{table}[h]
\begin{center}
\begin{tabular}{|c|c||r|r|r|r|}
\hline
expt. & \# of pts  & CTEQ3M & MRS A & CTEQ2M & MRS D-' \\ \hline
BCDMS$^{H}$ & 168     & 130.0(0.77)  & 168.0(1.00) & 110.2(0.66) & 133.2(0.79)
\\
BCDMS$^{D}$ & 156     & 187.2(1.20)  & 215.3(1.38) & 174.7(1.12) & 162.2(1.04)
\\
NMC$^H$ & 52          &  59.9(1.15)  &  60.2(1.15) &  61.5(1.18) &  59.2(1.14)
\\
NMC$^D$ & 52          &  47.2(0.91)  &  56.0(1.08) &  49.1(0.94) &  49.7(0.96)
\\
NMC$_{R}$ & 89        & 133.5(1.50)  & 140.6(1.58) & 139.7(1.57) & 144.2(1.62)
\\
CCFR $F_2$ & 63       &  69.3(1.10)  &  68.7(1.09) &  58.8(0.93) &  95.8(1.52)
\\
CCFR $F_3$ & 63       &  41.0(0.65)  &  61.7(0.98) &  37.2(0.59) &  67.4(1.07)
\\
ZEUS & 56  	      &  27.9(0.50)  &  40.3(0.72) &  27.9(0.50) &  74.5(1.33) \\
H1 & 21  	      &   7.7(0.37)  &   7.0(0.33) &   6.4(0.30) &  11.7(0.56) \\
E605 & 119 	      &  92.3(0.78)  &  95.9(0.81) &  88.1(0.74) & 102.6(0.86) \\
CDF DY & 8   	    &   3.0(0.38)  &   1.4(0.18) &   2.6(0.32) &   2.8(0.34) \\
CDF A$_{W}$ & 9       &   3.5(0.39)  &   3.4(0.38) &  12.2(1.36) &   3.8(0.42)
\\
NA51 A$_{DY}$ & 1     &   0.4(0.35)  &   0.0(0.03) &   3.0(3.02) &  10.3(10.3)
\\
WA70 & 39  	      &  23.3(0.60)  &  21.3(0.55) &  22.6(0.58) &  21.4(0.55) \\
E706 & 8   	      &  11.8(1.47)  &  11.2(1.40) &  12.2(1.52) &  11.3(1.41) \\
UA6$^{\bar{P}P}$ & 8  &   1.8(0.23)  &   1.6(0.20) &   2.2(0.27) &   1.5(0.19)
\\
UA6$^{PP}$ & 8        &   6.8(0.85)  &   6.8(0.85) &   7.5(0.94) &   6.8(0.85)
\\ \hline
Total & 920           &  844 & 959 & 816 & 958 \\ \hline
\end{tabular}
\caption{
$\chi^2$ and $\chi^2$ per point (in paranthesis) in each experiment and overall
for
current and previous version of CTEQ and MRS distributions. In the case of the
MRS distributions, we have {\em minimized} the $\chi ^2$
by adjusting all the experimental normalizations {\em freely}
while keeping the parton distributions as given by the authors.
(See Table \protect \ref{tbl:ExpNorm}.) These $\chi
^2$ values are obtained using the data sets of Table
\protect\ref{tbl:dataset}, employing the same error definitions
(except for experimental normalization for which the CTEQ numbers include
extra $\chi^2$'s for any deviation away from unity as explained in the text);
hence
they are not necessarily the same
as those quoted in the original work which may use a different selection of
data
points (e.g. for Drell-Yan, and direct photon experiments), apply different
error
definitions, and adopt different analysis procedures. Large differences
in the total $\chi^2$ are mainly associated with the precise BCDMS and CCFR
experiments. They may be partially attributed to the influence on the fits due
to the
$x<0.09$ data points of CCFR and NMC which are excluded in the CTEQ analyses
for consistency considerations, but included in the MRS ones.}
 \label{tbl:chisq}
\end{center}
\end{table}
}

\def\tblExpNorm
{
\begin{table}[h]
\begin{center}
\begin{tabular}{|l||c|c|c|c|}
\hline
expt. &              CTEQ3M & MRS A & CTEQ2M & MRS D-' \\
\hline
BCDMS &	             0.988 & 0.977 & 0.988 & 0.969 \\
NMC$_{90}$ &         1.008 & 1.008 & 1.009 & 0.996 \\
NMC$_{280}$ &        1.021 & 1.014 & 1.021 & 1.000 \\
CCFR &               0.975 & 0.968 & 0.979 & 0.958 \\
ZEUS &               0.978 & 1.029 & 1.003 & 1.061 \\
H1 &                 0.966 & 0.978 & 0.956 & 1.043 \\
E605 &               1.098 & 1.008 & 1.063 & 1.052 \\
CDF DY &             0.965 & 0.805 & 0.971 & 0.970 \\
WA70 &               1.010 & 1.055 & 1.023 & 0.977 \\
E706 &               0.923 & 0.980 & 0.960 & 0.912 \\
UA6$^{\bar{P}P}$ &   0.858 & 0.853 & 0.878 & 0.813 \\
UA6$^{PP}$ &         0.853 & 0.900 & 0.892 & 0.834 \\
\hline
\end{tabular}
\caption{
Normalization factors for experiments obtain in the CTEQ fits according to
error treatment procedure described in the text. For comparison, also
listed are normalization factors obtained by fitting the same data sets using
the
(fixed) MRS distributions and allowing all the experimental normalizations to
adjust {\em freely}. }
\label{tbl:ExpNorm}
\end{center}
\end{table}
}

\newpage
\tbldataset
\tblparamtwo
\tblparamthree
\tblchisq
\tblExpNorm

\end{document}